\documentclass[reprint,twocolumn,notitlepage,preprintnumbers,nofootinbib,amsmath,amssymb,aps,prd,floatfix,superscriptaddress,longbibliography,svgnames]{revtex4-1}

\usepackage{amsmath}
\usepackage{amsfonts}
\usepackage{amssymb}
\usepackage{graphicx}
\usepackage{bm}
\usepackage{subfigure}
\usepackage{url}
\usepackage[hyperindex]{hyperref}
\usepackage{color}
\usepackage{lineno}
\usepackage[ddmmyy,24hr]{datetime}
\usepackage{bigdelim}
\usepackage{booktabs}
\usepackage{dcolumn}
\usepackage{multirow}
\usepackage{subfigure}
\usepackage{physics}
\usepackage{cancel}
\usepackage{stackrel}
\usepackage{paralist}
\usepackage{xspace}
\usepackage{slashed}
\usepackage{cancel}
\usepackage{todonotes}
\usepackage{enumerate}
\usepackage{float}
\usepackage{fullpage}
\usepackage{ulem}
\usepackage{wasysym}
\usepackage{comment}
\usepackage{bbold}
\usepackage{hyperref}

\usepackage{xcolor}
\usepackage{orcidlink}

\usepackage{stackengine}

\usepackage{feynmp-auto}
\newcommand{\nua}[1]{\ensuremath{\rlap{\kern-2.5pt\ensuremath{\overset{\scriptscriptstyle(-)}{\phantom{\nu}}}}{\ensuremath{{\nu}_{#1}}}}}

\usepackage{enumitem}

\begin{document}

\title{A consistent neutrino determination of the low-energy weak mixing angle}

\author{M. Atzori Corona~\orcidlink{0000-0001-5092-3602}}
\email{mcorona@roma2.infn.it}
\affiliation{Istituto Nazionale di Fisica Nucleare (INFN), Sezione di Roma Tor Vergata, Via della Ricerca Scientifica, I-00133 Rome, Italy}

\author{M. Cadeddu~\orcidlink{0000-0002-3974-1995}}
\email{matteo.cadeddu@ca.infn.it}
\affiliation{Istituto Nazionale di Fisica Nucleare (INFN), Sezione di Cagliari,
	Complesso Universitario di Monserrato - S.P. per Sestu Km 0.700,
	09042 Monserrato (Cagliari), Italy}

\author{N. Cargioli~\orcidlink{0000-0002-6515-5850}}
\email{nicola.cargioli@ca.infn.it}
\affiliation{Istituto Nazionale di Fisica Nucleare (INFN), Sezione di Cagliari,
	Complesso Universitario di Monserrato - S.P. per Sestu Km 0.700,
	09042 Monserrato (Cagliari), Italy}

\author{F.\,Dordei~\orcidlink{0000-0002-2571-5067}}
\email{francesca.dordei@cern.ch}
\affiliation{Istituto Nazionale di Fisica Nucleare (INFN), Sezione di Cagliari,
	Complesso Universitario di Monserrato - S.P. per Sestu Km 0.700,
	09042 Monserrato (Cagliari), Italy}

\author{L. Ferro \orcidlink{0009-0002-1698-3710}}
\email{luca.ferro@ca.infn.it}
\affiliation{Istituto Nazionale di Fisica Nucleare (INFN), Sezione di Cagliari,
	Complesso Universitario di Monserrato - S.P. per Sestu Km 0.700,
	09042 Monserrato (Cagliari), Italy}
\affiliation{Dipartimento di Fisica, Universit\`{a} degli Studi di Cagliari,
	Complesso Universitario di Monserrato - S.P. per Sestu Km 0.700,
	09042 Monserrato (Cagliari), Italy}

\author{C. Giunti~\orcidlink{0000-0003-2281-4788}}
\email{carlo.giunti@to.infn.it}
\affiliation{Istituto Nazionale di Fisica Nucleare (INFN), Sezione di Torino, Via P. Giuria 1, I--10125 Torino, Italy}

\author{R. Pavarani~\orcidlink{0009-0004-9534-542X}}
\email{riccardo.pavarani@ca.infn.it}
\affiliation{Istituto Nazionale di Fisica Nucleare (INFN), Sezione di Cagliari,
	Complesso Universitario di Monserrato - S.P. per Sestu Km 0.700,
	09042 Monserrato (Cagliari), Italy}
\affiliation{Università degli Studi di Padova, Dipartimento di Fisica e Astronomia “Galileo Galilei”, Via Francesco Marzolo 8, 35131 Padova, Italy}

\author{M. Sestu~\orcidlink{0009-0002-6044-7800
}}
\email{michela.sestu@ca.infn.it}
\affiliation{Istituto Nazionale di Fisica Nucleare (INFN), Sezione di Cagliari,
	Complesso Universitario di Monserrato - S.P. per Sestu Km 0.700,
	09042 Monserrato (Cagliari), Italy}
\affiliation{Dipartimento di Fisica, Universit\`{a} degli Studi di Cagliari,
	Complesso Universitario di Monserrato - S.P. per Sestu Km 0.700,
	09042 Monserrato (Cagliari), Italy}

\begin{abstract}
The weak mixing angle, $\sin^2\theta_W$, is a fundamental parameter of the Standard Model and a sensitive probe of electroweak physics across various energy scales. At low momentum transfer, the Standard Model prediction approaches a well-defined low-energy plateau, providing a target for precision tests of the electroweak theory. At the lowest momentum transfer, atomic parity violation (APV) in cesium provides one of the most precise determinations of the weak mixing angle. Here, we review and update the status of this determination and compare it with other available probes. We then consistently reanalyze low-energy neutrino scattering data and combine the resulting determinations from coherent elastic scattering (CE$\nu$NS) and neutrino–electron scattering. The complementary sensitivity of these processes to electroweak couplings enables a neutrino-based determination of the low-energy weak mixing angle, accounting for electroweak radiative corrections and relevant theoretical uncertainties. Our global analysis yields $\sin^2\theta_W^{\rm low}(\nu\,\text{data}) = 0.232\pm0.006$.
We assess the consistency and relative impact of the individual neutrino measurements and discuss the prospects for future CE$\nu$NS and neutrino–electron scattering experiments to improve the precision of this determination. Our analysis provides an independent test of the Standard Model at low momentum transfer, complementary to that provided by atomic parity violation.
\end{abstract}

\maketitle  
\section{Introduction}
The weak mixing angle, $\sin^2\theta_W$, is one of the fundamental parameters of the Standard Model (SM) of electroweak interactions~\cite{Weinberg:1967tq}. Beyond tree level, it becomes scale dependent through electroweak radiative corrections, and its running with momentum transfer provides a sensitive test of the quantum structure of the electroweak theory~\cite{Erler:2004cx,Erler:2004in,Erler:2013xha,Erler:2017knj}. Precision measurements at different energy scales therefore probe not only the value of the weak mixing angle, but also the consistency of its predicted evolution~\cite{ParticleDataGroup:2024cfk,Erler:2026lyv,Kumar:2013yoa,ALEPH:2005ab}.\\
At low momentum transfer, the SM prediction approaches a well-defined plateau~\cite{Erler:2017knj,Erler:2024lds}. This low-energy regime provides a particularly interesting target for precision measurements: different observables probe essentially the same electroweak quantity while relying on very different experimental techniques and theoretical inputs. A consistent determination from independent low-energy probes therefore provides a non-trivial test of the SM and of its predicted running.

Atomic parity violation (APV), particularly in cesium, provides one of the most precise
determinations of the low-energy weak mixing angle~\cite{Roberts:2014bka,BOUCHIAT198373}. Its interpretation
requires accurate atomic-structure calculations and nuclear inputs,
whose refinement has affected the extracted value over time.
This illustrates the importance of independent low-energy observables
with different experimental and theoretical uncertainties.

Neutrino scattering provides a qualitatively different probe of the same
low-energy electroweak physics. In particular, coherent elastic
neutrino--nucleus scattering (CE$\nu$NS) and neutrino--electron scattering ($\nu$ES)
probe complementary combinations of the electroweak couplings.
Their combination therefore offers a neutrino-based determination of
the weak mixing angle that is independent of the atomic-structure
calculations entering APV. The complementarity of these low-energy
probes has also been emphasized in recent reviews of precision
electroweak physics below the $Z$ pole~\cite{Erler:2026lyv}.

In this work, we first review and update the determination of the weak
mixing angle from APV in cesium, accounting for recent atomic and nuclear
inputs. We then consistently reanalyze available low-energy CE$\nu$NS
and $\nu$ES measurements within a common
electroweak framework and combine their constraints into a neutrino-only
determination. We compare this result with the updated APV determination,
other parity-violating probes, and the SM prediction.  Throughout this work, $\sin^2\theta_W^{\rm low}$ denotes the
$\overline{\mathrm{MS}}$ weak mixing angle in the zero-momentum-transfer
limit. Its extraction includes the momentum-transfer dependence of the
electroweak radiative corrections relevant to each dataset.

\section{Atomic parity violation status}\label{sec:APV}
Measurements of parity-violating transitions in $^{133}\mathrm{Cs}$ have played a central role in establishing the sensitivity of atomic observables to the weak neutral-current interaction~\cite{Dzuba:2012kx, Bennett:1999pd,PhysRevA.65.012106,Viatkina:2019wsz,Ginges:2003qt,PhysRevC.46.2587,James_1999,Pollock1999,Horowitz2001}. The measured parity-violating amplitude is related to the weak charge of the nucleus, which at leading order~(LO) can be written as
\begin{equation}
Q^{\rm LO}_W \simeq -N+Z\left(1-4\sin^2\theta_W\right),
\end{equation}
where $N$ and $Z$ denote the neutron and proton numbers of the nucleus, respectively. At the precision of modern measurements, electroweak radiative corrections become important and, when taken into account, the weak charge can be expressed as~\cite{ParticleDataGroup:2024cfk}
\begin{align}
Q_W^{\mathrm{th}}=&- 2 [ Z (g_{A V}^{e p}(\sin^2\theta_W) + 0.00005)\nonumber\\
&+ N (g_{A V}^{e n} + 0.00006) ] \left( 1 - \dfrac{\alpha}{2 \pi} \right),
\label{QWSMthe}
\end{align}
where $\alpha$ is the fine-structure constant and the couplings of electrons to nucleons,
$g_{A V}^{e p}$ and $g_{A V}^{e n}$, include the effect of radiative corrections~\cite{ParticleDataGroup:2024cfk,Erler:2013xha,reverler,Cadeddu:2021ijh,AtzoriCorona:2023ktl} and lead to a SM value of the $^{133}\rm{Cs}$ weak charge of \mbox{$Q_{{W}}^{\text{SM}}=-73.27 \pm 0.01$}, when adopting \mbox{$\sin^2\theta_W^{\rm SM,\,low}=0.23873(5)$~\cite{ParticleDataGroup:2024cfk}} .\\
Experimentally, the weak charge of a nucleus is extracted from the ratio of the parity violating amplitude, $E_{\text{PNC}}$, to the Stark vector transition polarizability, $\beta$, and by calculating theoretically $E_{\rm PNC}$ in terms of $Q_W$
\begin{equation}
\label{QWeq}
Q^{\rm exp}_W\!=\! N \left( \dfrac{{\rm Im}\, E_{\rm PNC}}{\beta} \right)_{\rm exp.} \!\!
\left( \dfrac{Q_W}{N\, {\rm Im}\, E_{\rm PNC}} \right)_{\rm th.} \!\!\!\beta_{\rm exp.+th.},
\end{equation}
where $\beta_{\rm exp.+th.}$ can be determined semi-empirically by combining experimental measurements with precise theoretical calculations~\cite{ParticleDataGroup:2024cfk}, while $(\mathrm{Im}\, E_{\rm PNC})_{\rm th.}$ is calculated from atomic theory~\cite{Dzuba:2012kx,Sahoo:2021thl} and experimental inputs, and Im stands for the imaginary part.\\ 
Over the years, various values of $\beta_{\rm exp.+th.}$ were reported, and recently a new experimental average was derived~\cite{ParticleDataGroup:2024cfk}. 
The current PDG determination adopts the calculation of Ref.~\cite{Dzuba:2012kx}, while the alternative calculation of Ref.~\cite{Sahoo:2021thl} leads to a non-negligible shift in the extracted weak mixing angle, as extensively discussed in Refs.~\cite{AtzoriCorona:2023ktl,Sahoo:2021thl}.
Furthermore, nuclear effects, including the spatial distribution of neutrons, also enter at the level relevant for precision determinations~\cite{Viatkina:2019wsz,James_1999,Corona:2021yfd,PhysRevC.46.2587, Pollock1999,Horowitz2001,PhysRevA.65.012106,AtzoriCorona:2024vhj}. 
In particular, the so-called ``neutron-skin" correction was introduced in Ref.~\cite{PhysRevA.65.012106} to account for the difference between the neutron and proton root-mean-square radii, $R_n$ and $R_p$, respectively, which is neglected in the nominal atomic-theory calculation.
The latter has also been reconsidered and discussed in detail in more recent works~\cite{Cadeddu:2018izq,Cadeddu:2021ijh,AtzoriCorona:2023ktl}.
The neutron-skin-corrected weak charge is obtained by first removing
the neutron-skin correction included in the reference atomic calculation
and then adding to the resulting theoretical amplitude the updated
correction term~\cite{Viatkina:2019wsz,Cadeddu:2018izq}
\begin{equation}
\delta E_\mathrm{PNC}(R_n)= \left[ \frac{\mathrm{N}}{Q_W}\left(1-\frac{q_n(R_n)}{q_p}\right) E_\mathrm{PNC} \right].
\end{equation}
The factors $q_p$ and $q_n$ incorporate the radial dependence of the electron axial transition matrix element, $f(r)$, by considering the proton and neutron spatial distributions, $\rho_{p} (r)$ and $\rho_{n} (r)$, respectively~\cite{PhysRevC.46.2587,Pollock1999,James_1999,Horowitz2001,Viatkina:2019wsz}, 
\begin{equation}
    q_{p,n}=4\pi \int_0^\infty \rho_{p,n}(r) f(r) r^2\mathrm{d}r.
    \label{eq:qpn}
\end{equation}
The calculation of $q_p$ and $q_n$ has been extensively described in the Supplemental Material of Ref.~\cite{Cadeddu:2021ijh}, by considering a power expansion in $(Z\alpha)$, with $Z$ being the proton number. 
In this work, we update this calculation using an improved treatment of electron wave functions by numerically solving the Dirac-Hartree-Fock-Slater equations in a realistic atomic potential using the \text{RADIAL} package~\cite{radial}. 
The radial dependence of the electron axial transition matrix element,
$f(r)$, is then evaluated from the corresponding $s_{1/2}$ and $p_{1/2}$
Dirac wave functions inside the nucleus and normalized to $f(0)=1$.\\
Moreover, the neutron skin correction depends crucially on the employed value for the neutron radius of cesium. 
The PDG determination adopts the neutron-skin correction associated with the atomic calculation of Ref.~\cite{Dzuba:2012kx}, using the fixed value $R_n(^{133}\rm{Cs})\simeq4.93~\mathrm{fm}$ obtained from an extrapolation of antiprotonic-atom x-ray data~\cite{Trzcinska:2001sy}, without propagating the uncertainty on the neutron radius. 
The possibility of obtaining a fully data-driven simultaneous determination of the weak mixing angle and the neutron skin by combining APV with CE$\nu$NS data has also been explored in Refs.~\cite{Cadeddu:2021ijh,AtzoriCorona:2023ktl}.
Here, we consider the neutron skin determination obtained by exploiting proton-cesium elastic scattering as reported in Ref.~\cite{Huang:2024jbh} and discussed in Ref.~\cite{AtzoriCorona:2024vhj}, to determine the experimental neutron radius, $R_n(^{133}\rm{Cs})=4.94\pm0.21~\mathrm{fm}$. Instead, for the nuclear proton radius, we correct the experimental charge radius determined from muonic atom spectroscopy~\cite{Fricke:1995zz,Angeli:2013epw} as in Ref.~\cite{Cadeddu:2020lky}, obtaining
$R_{p}(^{133}\text{Cs})=4.821 \pm 0.005 \, \text{fm}$.\\ 
Our refined suggested values are
\begin{equation}
    q_n=0.953\pm0.003\, ,\quad q_p=0.95522\pm0.00008\, ,
\end{equation}
corresponding to relative differences of $-0.19\%$ and $-0.15\%$, respectively, compared to the values employed in Ref.~\cite{Cadeddu:2021ijh}.\\
Considering these inputs, the latest determinations from APV on cesium result in 
\begin{align}
    \sin^2\theta_W^{\rm low}(\text{APV})=
    \begin{cases}   
    0.2348\pm0.0022\, ,\\
    0.2372\pm0.0020\, ,\end{cases}
\end{align}
when using the atomic calculation from Ref.~\cite{Dzuba:2012kx} and that from Ref.~\cite{Sahoo:2021thl}, respectively.
The reported uncertainties are slightly larger with respect to previous determinations. This reflects the incorporation of the experimental neutron skin uncertainty into the extracted weak mixing angle determination.\\
 \begin{figure}[h]
    \centering
    \includegraphics[width=\columnwidth]{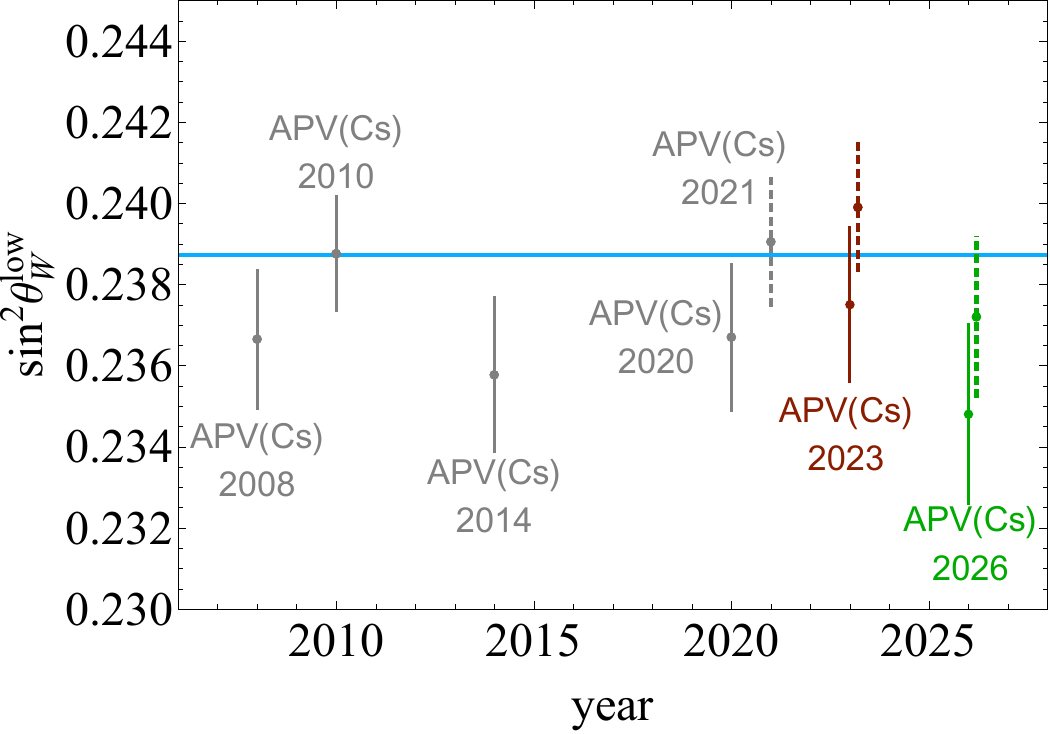}
    \caption{Summary of the weak mixing angle measurements from APV(Cs) across the years. The light blue line shows the SM value of the weak mixing angle at low energies. The gray points show the measurements from APV on cesium atoms during the years~\cite{ParticleDataGroup:2008zun,ParticleDataGroup:2010dbb,ParticleDataGroup:2014cgo,Zyla:2020zbs,ParticleDataGroup:2022pth,Sahoo:2021thl}. Our re-determinations of the 2020 and 2021 results are indicated by the dark-red points~\cite{AtzoriCorona:2023ktl}, while the latest values are shown by the green points. Dashed uncertainty lines refer to data obtained with the theoretical atomic calculation in Ref.~\cite{Sahoo:2021thl}, while solid uncertainty lines denote the data obtained by considering the reference value reported by the PDG~\cite{ParticleDataGroup:2026mpi}.}
    \label{fig:APV_history}
\end{figure}

The history of the cesium results provides an instructive example of the interplay between experimental precision and theoretical interpretation. The evolution of the value of $\sin^2\theta_W$ inferred from APV is shown in Fig.~\ref{fig:APV_history}. The high-precision cesium measurement initially yielded a deviation from the SM prediction when combined with the atomic calculations available at the time. Subsequent improvements in the treatment of atomic correlations and other corrections substantially reduced this discrepancy and brought the result into agreement with the SM. More recent calculations of the parity-nonconserving amplitude have led to further shifts in the extracted weak mixing angle. \\
The evolution shown in Fig.~\ref{fig:APV_history} should not be interpreted as a limitation of atomic parity violation. On the contrary, it illustrates the extraordinary sensitivity of APV to the low-energy electroweak interaction. It also demonstrates a general principle of precision tests: as experimental uncertainties become increasingly small, independent observables with different systematics, such as neutrino scattering, become essential. Agreement between such probes provides a more robust validation of the SM than any single determination considered in isolation. Thus, the purpose of the following analysis is not to improve upon the APV precision, but to provide an independent determination based solely on neutrino scattering data. 
For the comparison with the neutrino determination, we adopt as our updated APV benchmark
\mbox{$\sin^2\theta_W^{\rm low}(\text{APV})=0.2348\pm0.0022$},
obtained using the atomic calculation of Ref.~\cite{Dzuba:2012kx}.
This choice follows the atomic calculation adopted by the PDG~\cite{ParticleDataGroup:2026mpi}.
We also consider the alternative determination based on
Ref.~\cite{Sahoo:2021thl}.

\section{Low-energy neutrino scattering}
\subsection{Coherent elastic neutrino--nucleus scattering}
Coherent elastic neutrino--nucleus scattering is a process in which an entire nucleus ($\mathcal{N}$) is scattered coherently by a neutral-current interaction with a low-energy neutrino of flavor $\ell$. The coherence enhancement makes CE$\nu$NS particularly sensitive to the weak charge of the target and therefore to the weak neutral-current couplings.
\\
For a spin-zero nucleus, the SM differential cross section as a function of the nuclear recoil energy $T_\mathrm{nr}$ can be written as~\cite{Cadeddu:2023tkp}
\begin{equation}
    \frac{d\sigma_{\nu_{\ell}\text{-}\mathcal{N}}}{d T_\mathrm{nr}} = 
    \frac{G_{\text{F}}^2 M}{\pi} 
    \left( 1 - \frac{M T_\mathrm{nr}}{2 E_\nu^2} \right)
    \left( Q^{V}_{\nu_\ell, \mathrm{SM}} \right)^2,
    \label{eq:cexsec}
\end{equation}
where $G_{\text{F}}$ is the Fermi constant, $E_\nu$ the neutrino energy, $M$ the nuclear mass, and $ Q^{V}_{\nu_\ell, \mathrm{SM}}$ is the nuclear weak charge. Similarly to the case of APV in Eq.~(\ref{QWSMthe}), the nuclear weak charge can be expressed in terms of the proton and neutron couplings, $g_V^p$ and $g_V^n$, namely 
\begin{equation}
Q^{V}_{\nu_{\ell}, \mathrm{SM}}=
g_V^p Z F_p(q^2)
+
g_V^n N F_n(q^2),
\end{equation}
with $F_{p,n}(q^2)$ being the proton and neutron form factors.\\
The dependence on the weak mixing angle is hidden inside the proton coupling. However, since the neutron contribution dominates the total weak charge, the dependence on $\sin^2\theta_W$ is relatively weak compared with the overall normalization of the process, making precise control of flux and detector normalization particularly important.\\
At finite momentum transfer, the nuclear form factors are sensitive to the neutron and proton distributions. In particular, uncertainties in the neutron distribution can become relevant as the experimental precision improves~\cite{Sierra:2023pnf,AbdelKhaleq:2024hir,AtzoriCorona:2023ktl}. This introduces a nuclear-structure component into the extraction of $\sin^2\theta_W$, which is nevertheless qualitatively distinct from the atomic-structure uncertainties entering APV.

\subsection{Neutrino--electron scattering}
Neutrino--electron scattering provides a complementary probe of the weak neutral-current interaction. Unlike CE$\nu$NS, the target is an elementary electron and no nuclear-structure uncertainty enters the cross section. The latter can be written as~\cite{ParticleDataGroup:2024cfk,Kouzakov:2017hbc}
\begin{align}\label{ES-SM-flavour}
\dfrac{d\sigma_{\nu_{\ell}\,e}}{d T_{\text{e}}}&
%(E_\nu,T_{e})
=
Z_{\text{eff}}(T_{e})
\,
\dfrac{G_{\text{F}}^2 m_{e}}{2\pi}
\left[
\left( g_{V}^{\nu_{\ell}\,e} + g_{A}^{\nu_{\ell}\,e} \right)^2\right.\nonumber\\\nonumber
&+
\left( g_{V}^{\nu_{\ell}\,e} - g_{A}^{\nu_{\ell}\,e} \right)^2
\left( 1 - \dfrac{T_{e}}{E_\nu} \right)^2\\
&-
\left.\left( \left(g_{V}^{\nu_{\ell}\,e}\right)^2 - \left(g_{A}^{\nu_{\ell}\,e}\right)^2 \right)
\dfrac{m_{e} T_{e}}{E_\nu^2}
\right]
,
\end{align}
where $m_{e}$ is the electron mass, $Z_{\rm eff}$ describes the effective number of electrons that can be ionized by an energy deposit $T_e$~\cite{Kouzakov:2017hbc} (for a scattering on a free electron,
$Z_{\rm eff}=1$), while $g_{V}^{\nu_{\ell}\,e}$ and $g_{A}^{\nu_{\ell}\,e}$, are the vector and axial-vector couplings. In particular, the vector coupling depends on $\sin^2\theta_W$. For antineutrinos, the cross section is obtained by interchanging
$g_V+g_A$ and $g_V-g_A$. Moreover, for electron neutrinos, charged-current and neutral-current contributions both enter, whereas for muon and tau neutrinos only the neutral-current contribution is present.\\
The different dependence on the electroweak couplings is important when computing the combined determination. Neutrino--electron scattering provides information that is complementary to CE$\nu$NS as it is affected by different experimental and theoretical uncertainties. In particular, it does not suffer from the neutron-distribution uncertainty associated with coherent scattering off nuclei.\\

The combination of CE$\nu$NS and $\nu$ES therefore provides a neutrino-only determination of the weak mixing angle based on two physically distinct classes of observables.
Moreover, throughout our analysis, the electroweak couplings are evaluated consistently within the same framework for the calculation of the electroweak radiative corrections~\cite{Erler:2013xha,AtzoriCorona:2024rtv,AtzoriCorona:2025xwr}.

\section{Global Fit of Neutrino Data}
\begin{figure}[t]
    \centering
    \includegraphics[width=\columnwidth]{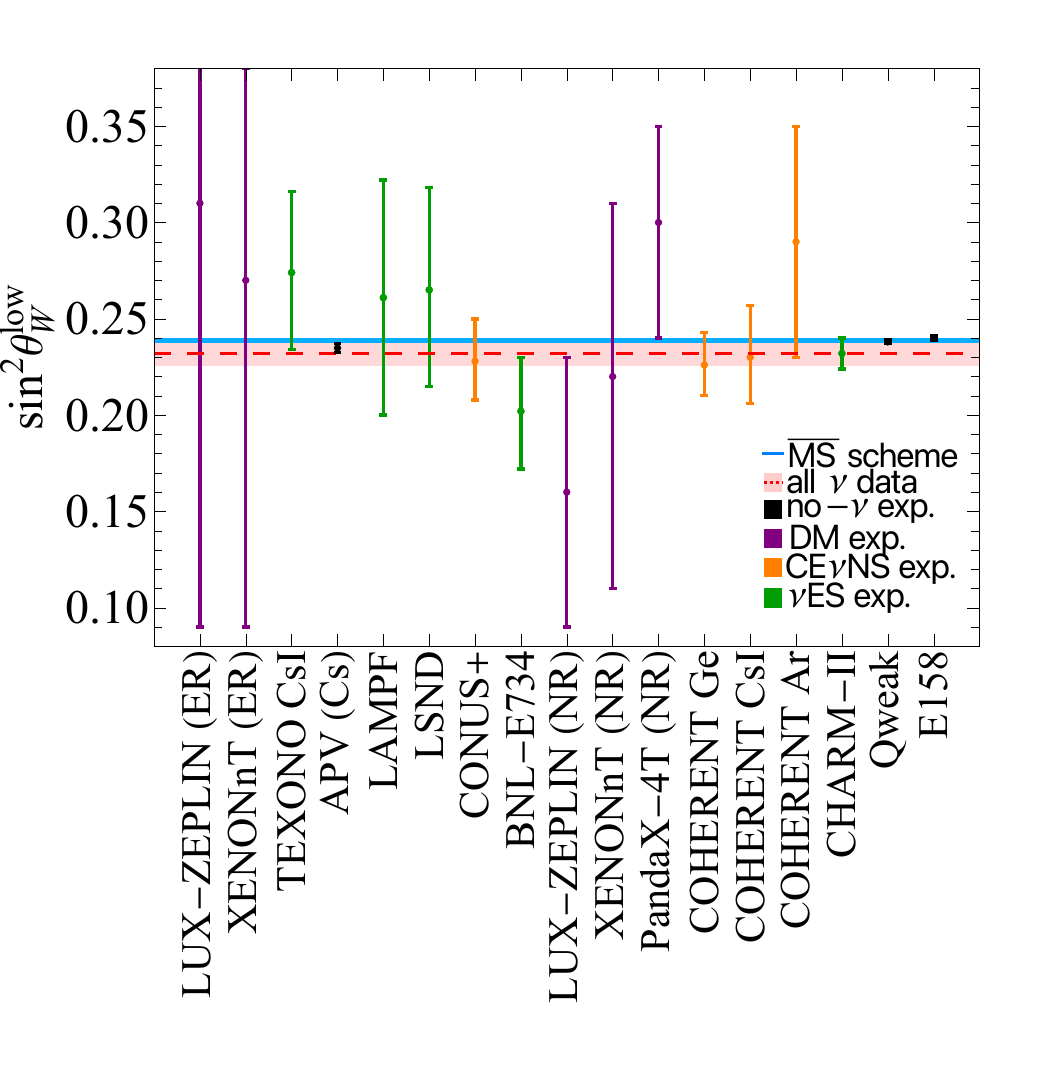}
    \caption{Low-energy weak mixing angle determinations obtained in this work for the various neutrino data as reported in Tab.~\ref{tab:Summary}, compared with the parity-violating probes (black), the SM running (blue) and the neutrino data global fit result (red).}
    \label{fig:nu_measurements}
\end{figure}

In this work we consistently re-determine, within a common low-energy electroweak framework, the values of $\sin^2\theta_W^{\rm low}$ obtained by the different experiments under consideration, including the reanalysis of older measurements for which no determination of the weak mixing angle was available or previous determinations relied on earlier theoretical inputs.
The resulting individual $\chi^2$ profiles are then combined into a global neutrino determination. \\ \\
Building upon the framework developed for the global fit of neutrino charge radii and neutral-current couplings in Ref.~\cite{AtzoriCorona:2025xwr}, here we extend the analysis to the extraction of the weak mixing angle by exploiting the large amount of available $\nu$ES and CE$\nu$NS data, including newly available datasets not considered in Ref.~\cite{AtzoriCorona:2025xwr}.
In particular, the global analysis includes the CE$\nu$NS data from COHERENT CsI~\cite{COHERENT:2021xmm}, analyzed following Ref.~\cite{AtzoriCorona:2023ktl}; COHERENT Ar~\cite{COHERENT:2020iec,COHERENT:2020ybo}, following Ref.~\cite{Cadeddu:2020lky}; the two COHERENT Ge data releases~\cite{COHERENT:2025vuz,COHERENT:2026yje}, analyzed in combination as described in Ref.~\cite{AtzoriCorona:2026wbu}.
Here, we consider the results from the first scientific run (Run 1) of the CONUS+ experiment~\cite{Ackermann:2025obx}, as in Ref.~\cite{AtzoriCorona:2025ygn}, in combination for the first time with the newly released CONUS+ high precision data (Run 2)~\cite{Ackermann:2026mms} following the details reported by the collaboration in Ref.~\cite{Ackermann:2026bup}. The collaboration reports a total of $1555\pm152$ events combining their three detectors, to be compared with a total predicted number of events of $1701\pm198$ in the SM. Our theoretical prediction, accounting for the energy resolution (corrected for the effect of non-linearity) and the detector efficiency, is $1710\pm199$ events. We fit our theoretical prediction to the experimental number of events reported for each individual detector, considering a common nuisance parameter accounting for the $11.6\%$ systematic uncertainty, obtaining $\sin^2\theta_W^{\rm low}(\rm CONUS+~Run~2)=0.222^{+0.023}_{-0.021}$, which is signficantly more precise than that reported in the previous data campaign~\cite{Ackermann:2025obx} and in excellent agreement to the official value reported by CONUS+~\cite{Ackermann:2026mms}. We also include the constraints from the recent TEXONO Ge data~\cite{TEXONO:2026eyr}, analyzed using the same strategy as in Ref.~\cite{AtzoriCorona:2025ygn}.
We also update $\nu$GeN constraints using the latest data~\cite{nGeN:2025hsd}. Moreover, we analyze the reactor $\overline{\nu}_e-e$ data from the TEXONO collaboration~\cite{TEXONO:2009knm} using a recent antineutrino flux parametrization~\cite{Perisse:2023efm}. We also consider the results on the $\nu_e-e$ integrated cross section measured by the LSND~\cite{LSND:2001akn} and LAMPF~\cite{Allen:1992qe} collaborations.
Similarly, we analyze the integrated cross section results reported by BNL-E734 for the $\nu_\mu (\overline{\nu}_\mu)-e$ processes~\cite{Ahrens:1990fp}.
We also include the $\nu_\mu (\overline{\nu}_\mu)-e$ differential cross section measurement reported by the CHARM-II experiment~\cite{CHARM-II:1993phx,CHARM-II:1994dzw,CHARM-II:1994aeb}.
\begin{table*}[t!]
\centering
\resizebox{1.6\columnwidth}{!}{
    \renewcommand{\arraystretch}{1.1}
\begin{tabular}{llllll}
\hline
\textbf{Experiment/}\hspace{0.1cm} & \multirow{2}{*}{\textbf{Channel}}\hspace{0.1cm} & \multirow{2}{*}{\textbf{Target}}\hspace{0.3cm}
& \multirow{2}{*}{\textbf{Source}}\hspace{0.6cm} & \textbf{Scale\ $Q$} & \multirow{2}{*}{$\sin^2\theta^\text{low}_W$}\\
\textbf{class} & & & & \textbf{[MeV]} \\
\hline

P4T (ER)~\cite{PandaX:2024cic}
    & $\nu-e$
    & $e$ in Xe
    & Solar $pp$
    & $\lesssim 1$ 
    &$<0.43 (1\sigma)$\\

LZ (ER)~\cite{LZ:2024zvo}
    & $\nu-e$
    & $e$ in Xe
    & Solar $pp$
    & $\lesssim 1$ 
    &$0.31^{+0.12}_{-0.22}$\\

XnT (ER)~\cite{XENON:2022ltv}
    & $\nu-e$
    & $e$ in Xe
    & Solar $pp$
    & $\lesssim 1$ 
    &$0.27^{+0.11}_{-0.18}$\\

TEXONO~\cite{TEXONO:2009knm}
    & $\bar{\nu}_e-e$
    & $e$ in CsI(Tl)
    & Reactor $\bar{\nu}_e$
    & $2$--$3$ 
    &$0.274^{+0.042}_{-0.040}$\\

\textbf{APV(Cs)}~\cite{ParticleDataGroup:2024cfk}
    & APV
    & Cs atom
    & Atomic
    & $2.4$ 
    &$0.2348\pm 0.0022$\tablenotemark[4]\\

LAMPF~\cite{Allen:1992qe}
    & $\nu_e-e$
    & $e$
    & $\mu$-DAR\,\tablenotemark[1]
    & $3$--$7$ 
    &$0.261\pm 0.061$\\
    
LSND~\cite{LSND:2001akn}
    & $\nu_e-e$
    & $e$
    & $\mu$-DAR\,\tablenotemark[1]
    & $4$--$7$ 
    &$0.265^{+0.053}_{-0.050}$\\

CONUS+~\cite{Ackermann:2025obx,Ackermann:2026mms}
    & CE$\nu$NS
    & Ge
    & Reactor $\bar{\nu}_e$
    & $5$--$20$ 
    &$0.228^{+0.022}_{-0.020}$\\

TEXONO~\cite{TEXONO:2026eyr}
    & CE$\nu$NS
    & Ge
    & Reactor $\bar{\nu}_e$
    & $5$--$20$ 
    &$<0.31 (1\sigma)$\\

$\nu$GeN~\cite{nGeN:2025hsd}
    & CE$\nu$NS
    & Ge
    & Reactor $\bar{\nu}_e$
    & $5$--$20$ 
    &$0.33^{+0.15}_{-0.25}$\\

BNL-E734~\cite{Ahrens:1990fp}
    & $\nu_\mu-e$
    & $e$
    & Accelerator
    & $10$--$40$ 
    &$0.202^{+0.028}_{-0.030}$\\

LZ (NR)~\cite{LZ:2025igz}
    & CE$\nu$NS
    & Xe
    & Solar ${}^{8}$B
    & $10$--$50$ 
    &$0.16\pm0.07$\\%\tablenotemark[4]\\

XnT (NR)~\cite{XENON:2024ijk}
    & CE$\nu$NS
    & Xe
    & Solar ${}^{8}$B
    & $10$--$50$ 
    &$0.22^{+0.09}_{-0.11}$\\

P4T (NR)~\cite{PandaX:2024muv}
    & CE$\nu$NS
    & Xe
    & Solar ${}^{8}$B
    & $10$--$50$ 
    &$0.30^{+0.05}_{-0.06}$\\    

COHERENT~\cite{COHERENT:2025vuz,COHERENT:2026yje}
    & CE$\nu$NS
    & Ge
    & $\pi$-DAR\,\tablenotemark[1]
    & $20$--$50$ 
    &$0.226^{+0.017}_{-0.016}$\\

COHERENT~\cite{COHERENT:2021xmm}
    & CE$\nu$NS
    & CsI
    & $\pi$-DAR\,\tablenotemark[1]
    & $30$--$60$ 
    &$0.230^{+0.027}_{-0.024}$\\

COHERENT~\cite{COHERENT:2020iec,COHERENT:2020ybo}
    & CE$\nu$NS
    & Ar
    & $\pi$-DAR\,\tablenotemark[1]
    & $20$--$100$ 
    &$0.29\pm0.06$\\

CHARM-II~\cite{CHARM-II:1993phx,CHARM-II:1994dzw,CHARM-II:1994aeb}
    & $\nu_\mu-e$
    & $e$
    & Accelerator
    & $140$ 
    &$0.232\pm0.008$\\

\textbf{Qweak}~\cite{Androic:2018kni}
    & $e-p$\,\tablenotemark[2]
    & $p$
    & $e$-beam\,\tablenotemark[3]
    & $160$ 
    &$0.238\pm 0.001$\\

\textbf{E158}~\cite{Anthony:2005pm}
    & M{\o}ller\,\tablenotemark[2]
    & $e$
    & $e$-beam\,\tablenotemark[3]
    & $161$ 
    &$0.240\pm0.001$\\

\hline
\end{tabular}
} 

\tablenotetext[1]{Decay-at-rest.}
\tablenotetext[2]{Parity violating channel.}
\tablenotetext[3]{Considering a polarized electron beam.}
\tablenotetext[4]{This represents the updated value obtained in this work when considering the atomic calculation from Ref.~\cite{Dzuba:2012kx}, which shows good agreement with the value reported in Ref.~\cite{ParticleDataGroup:2024cfk}.}
\caption{Selected low-energy parity-violation (bold) and neutrino-scattering
measurements, ordered approximately by characteristic momentum transfer
$Q$. The quoted $Q$ values are representative scales rather than fixed
event-by-event momentum transfers. The reported weak mixing angle values for neutrino data have either been obtained by us in previous publications or re-determined in the present analysis. }\label{tab:Summary}
\end{table*}
%%%%%%%%%%%%%%%%%%%%%%%%%%%%%%%%%%%
Furthermore, we incorporate neutrino-scattering constraints obtained from dark-matter direct-detection experiments, such as XENONnT (XnT)~\cite{XENON:2024ijk, XENON:2022ltv}, PandaX-4T~\cite{PandaX:2024muv, PandaX:2024cic} and LUX-ZEPLIN (LZ)~\cite{LZ:2024zvo,LZ:2025igz}, as described in Ref.~\cite{AtzoriCorona:2025gyz}.\\
For each experiment, we construct a chi-square function in terms of $\sin^2\theta_W$, incorporating the experimental uncertainties and the relevant nuisance parameters. Depending on the dataset, these include, in general, uncertainties associated with neutrino fluxes, detector response, normalization, backgrounds, quenching factor models and nuclear structure effects.
We report the individual constraints from each of the considered datasets in Fig.~\ref{fig:nu_measurements}. 
The figure shows the individual determinations entering the neutrino global fit, illustrating their relative constraining power and compatibility with the SM expectation. 
The data points are reported from the lower to the higher momentum transfer. In 
Tab.~\ref{tab:Summary}, we summarize the value of the weak mixing angle for each experimental result along with the channel, target, source and estimated momentum transfer. The neutrino results are compared to the best available non-neutrino determinations in the low-energy regime, namely the APV result discussed previously, and the electron and proton weak charges as measured by the E158~\cite{Anthony:2005pm} and Qweak~\cite{Androic:2018kni} collaborations, respectively.

The individual measurements are broadly compatible with each other and with the SM expectation, as illustrated in Fig.~\ref{fig:nu_measurements}. Neither CE$\nu$NS nor $\nu$ES alone currently provides a determination competitive with the APV result. 
Assuming the individual datasets to be statistically independent, the global $\chi^2$ profile is obtained by summing the corresponding profiled $\chi^2$ functions. 
This procedure naturally preserves asymmetric uncertainties and one-sided upper constraints, while allowing the impact of possible tensions among datasets to be reflected in the global profile.
The latter is then used to determine the best-fit value and confidence interval for $\sin^2\theta_W^{\rm low}$. \\
The combined analysis of neutrino data yields
\begin{equation}
\sin^2\theta_W^{\rm low}(\nu\,\text{data})=0.232\pm0.006,\label{resGF}
\end{equation}
and is shown in Fig.~\ref{fig:nu_measurements} by the red band, and lies slightly below the SM prediction \mbox{$\sin^2\theta_W^{\rm SM,\,low}=0.23873(5)$~\cite{ParticleDataGroup:2024cfk}} indicated by the blue line, 
whose value corresponds to a $\Delta\chi^2$ shift of $1.12$.
The current precision is limited by the combination of experimental statistics and systematic uncertainties in the available neutrino datasets. In particular, as shown by the $\nu-e$-only and CE$\nu$NS-only global fit reported in Tab.~\ref{tab:combined}, the $\nu$ES data are more precise than the CE$\nu$NS ones, although sharing approximately the same central value. Nevertheless, the result constitutes a direct and complementary determination of the low-energy weak mixing angle using observables that are independent of the atomic-structure calculations required for APV.
\begin{table}[h]
    \centering
    \resizebox{0.6\columnwidth}{!}{
    \renewcommand{\arraystretch}{1.1}
    \begin{tabular}{l|c}
       Data  &  $\sin^2\theta_W^{\rm low}$\\\hline
       all $\nu-e$ data  & $0.233\pm 0.008$\\
       all CE$\nu$NS data & $0.229\pm 0.011$\\
       all $\nu$ data & $0.232\pm0.006$\\\hline
       \end{tabular}}
    \caption{Results from our analysis of the available neutrino datasets decomposed by physical process and their combination.}
    \label{tab:combined}
\end{table}

\section{The neutrino determination in the global electroweak picture}

\begin{figure}[t]
    \centering
    \includegraphics[width=\columnwidth]{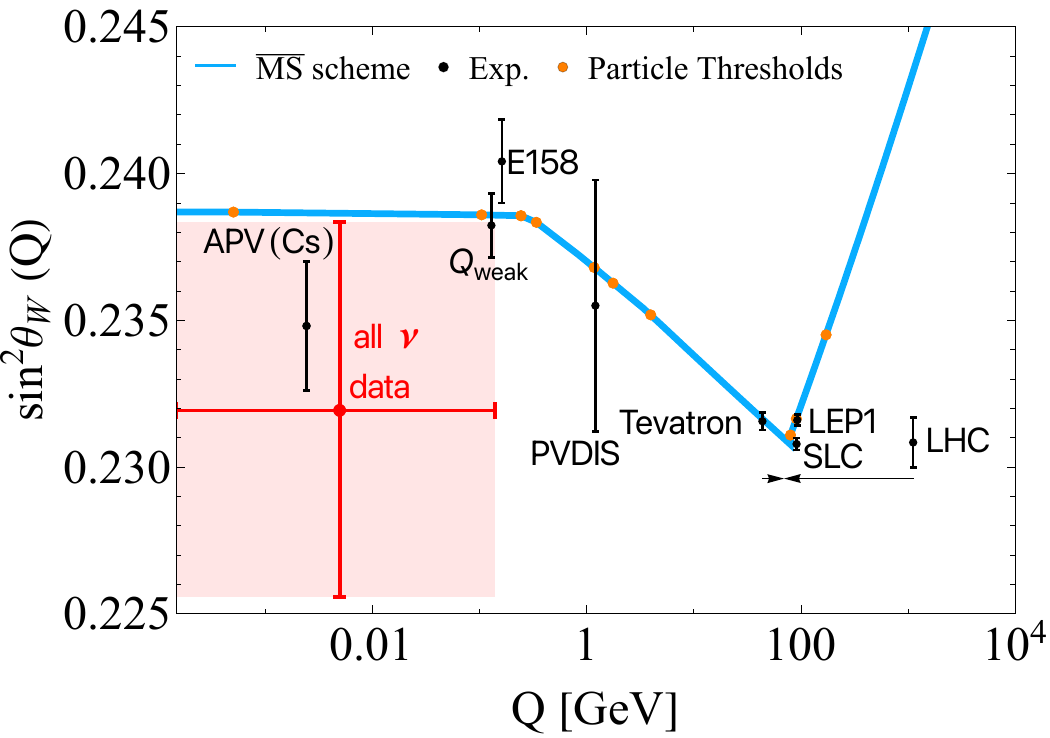}
    \caption{Running of the weak mixing angle with the energy scale as predicted by the SM in the $\overline{\mathrm{MS}}$ scheme~\cite{ParticleDataGroup:2026mpi}, compared to the available experimental determinations~\cite{Androic:2018kni,Anthony:2005pm,Wang:2014bba,ParticleDataGroup:2026mpi}. In red the result of the global fit of neutrino data obtained in this work. The red shaded region identifies the region constrained by the various neutrino data.}
    \label{fig:global_running}
\end{figure}

The result of the global neutrino fit is also reported in Fig.~\ref{fig:global_running} compared to the non-neutrino determinations across the energy scales. The running of $\sin^2\theta_W$ connects the high-energy electroweak scale with the low-momentum-transfer regime probed by neutrino scattering. At sufficiently low momentum transfer, the running approaches the low-energy plateau.
The neutrino measurements considered here probe precisely this energy regime. The combined value, reported in Eq.~(\ref{resGF}), lies below the value of the SM plateau, but remains compatible with it at the $\sim 1\sigma$ level given the present uncertainties.\\
The comparison with atomic parity violation is particularly informative. APV and neutrino scattering probe the same low-energy electroweak physics through fundamentally different observables. In APV, the weak interaction produces a parity-violating atomic transition whose interpretation requires atomic many-body calculations and nuclear corrections. In neutrino scattering, the weak interaction is probed directly through neutrino scattering, with CE$\nu$NS introducing nuclear-structure uncertainties of a different nature and $\nu$ES avoiding nuclear effects altogether.
\\
The two approaches should therefore be regarded as complementary rather than competing measurements. The higher precision currently achieved with APV makes it the sharper individual constraint, while the neutrino result provides an independent cross-check of the same low-energy electroweak prediction, although it is worth mentioning that the uncertainty of the combined neutrino determination is currently about three times that of the updated APV determination.

This independence is particularly relevant in light of the historical evolution of the APV determination discussed in Sec.~\ref{sec:APV}. A shift in the inferred value of $\sin^2\theta_W$ associated with improvements in the atomic or nuclear theory does not imply a change in the underlying experimental observable. An independent neutrino determination is therefore essential for establishing whether a future deviation from the SM is common to different low-energy observables or specific to the theoretical interpretation of a particular measurement.
This complementarity is particularly relevant in the broader low-energy electroweak program, where neutrino scattering, atomic parity violation, and parity-violating electron scattering provide independent probes of the same neutral-current dynamics~\cite{Erler:2026lyv}.
\\
The present neutrino result is consistent with the APV determination and with the SM running.
Considerable progress is still expected on the neutrino side, particularly from CE$\nu$NS, where a rapidly expanding experimental program offers significant potential for improved precision. Currently, a key limitation of the present COHERENT CE$\nu$NS constraints arises from the experimental systematics affecting the neutrino flux. To address this limitation, the COHERENT collaboration has developed a dedicated D$_2$O detector aimed at providing an in situ normalization of the SNS neutrino flux through neutrino--deuteron interactions~\cite{COHERENT:2021xhx}. Consequently, accelerator-based CE$\nu$NS experiments will become dominated by nuclear structure uncertainties, which, however, can be limited by combining measurements on the same nuclear target at different neutrino sources, e.g. accelerator+reactor experiments. Reactor experiments, instead, will be limited by the uncertainties in the detector response, including the quenching factor where relevant, and in the theoretical description of the reactor antineutrino flux. However, dedicated campaigns and efforts are planned within the community to better control these uncertainties. In the meantime, solar neutrino data are going to be accumulated, as dark matter detectors are expected to be running for years, pushing the statistics to the limit.
As the precision of such neutrino measurements improves, exploiting the complementarity of accelerator-, reactor-, and solar-neutrino probes~\cite{COHERENT:2026ewu,AtzoriCorona:2025ibl,NUCLEUS:2026pnv,CentellesChulia:2025jir,Maity:2024aji,Baxter:2019mcx,Chatterjee:2022mmu,Delgadillo:2026jee,Cao:2026nfm}, the comparison between these independent observables will become an increasingly stringent test of the low-energy electroweak theory.

\section{Conclusions}

We have presented a neutrino-based determination of the low-energy weak mixing angle, obtained by consistently reanalyzing a broad set of low-energy neutrino-scattering measurements and combining their individual constraints.
The result is consistent with the SM prediction for the low-energy plateau and with the available determinations from atomic parity violation in cesium. While the present combined neutrino constraint is less precise than APV, its significance lies in its independence from the atomic-structure calculations required to interpret the atomic measurement.
The historical evolution of the cesium APV determination illustrates why this independence is important. The exceptional precision of the APV measurement has been accompanied by continued improvements in atomic-structure calculations and nuclear inputs, which have affected the value inferred for the weak mixing angle. This history does not undermine the precision of APV; rather, it emphasizes the importance of testing the same Standard Model prediction through independent observables.
\\
Neutrino scattering provides such an independent probe. Neutrino--electron scattering directly tests the neutral-current couplings without nuclear-structure uncertainties, while CE$\nu$NS provides complementary sensitivity through coherent scattering from nuclei. Their combination therefore establishes a neutrino-based benchmark for low-energy weak mixing angle measurements.
\\
Neutrino scattering data are in good agreement with both APV and the Standard Model running, although the current comparison remains limited by experimental uncertainties.
Future CE$\nu$NS experiments are expected to improve this situation substantially, with percent-level sensitivities to $\sin^2\theta^{\rm low}_W$ projected for next-generation detectors. Together with improved neutrino--electron scattering measurements, these experiments can progressively close the precision gap with APV while maintaining their independent theoretical systematics.
The ultimate importance of a neutrino determination is therefore not simply to reproduce the precision of APV, but to establish the low-energy electroweak prediction through a qualitatively different set of observables. A future agreement between independent neutrino and atomic determinations at comparable precision would provide a stringent validation of the Standard Model running of the weak mixing angle. Conversely, a deviation consistently observed in both approaches would constitute a particularly robust indication of new physics.\\
The present global analysis of neutrino data represents a first step toward such a program, providing an independent low-energy benchmark and demonstrating the potential of neutrino scattering as a precision probe of the electroweak sector.

\bibliography{ref}

\end{document}